\documentclass[%
 reprint,
preprintnumbers,
 amsmath,amssymb,
 aps,
]{revtex4-2}

\usepackage{graphicx}
\usepackage{dcolumn}
\usepackage{bm}
\usepackage{hyperref}
\usepackage{xcolor}

\begin{document}

\preprint{MIT-CTP/6021}
\preprint{UUITP-05/26}

\title{3D gravity and double copy theory}

\author{Maor Ben-Shahar}
\email{maorbs@mit.edu}
\affiliation{MIT Center for theoretical physics - a Leinweber Institute, \\
Cambridge, MA 02139, USA}
\author{Francesco Bonechi}
\affiliation{INFN Sezione di Firenze, Via G. Sansone 1, \\
50019 Sesto Fiorentino, Firenze, Italy}
\email{francesco.bonechi@fi.infn.it}
\author{Maxim Zabzine}
\email{maxim.zabzine@physics.uu.se}
\affiliation{
Department of Physics and Astronomy, Uppsala University, \\
Box 516, SE-75120 Uppsala, Sweden
}
\affiliation{
Centre for Geometry and Physics, Uppsala University, \\
Box 480, SE-75106 Uppsala, Sweden
}

\date{\today}

\begin{abstract}
We introduce a novel reformulation of three-dimensional gravity in terms of divergenceless vector frames, inspired by the double copy for Chern-Simons theory. This formulation is on-shell equivalent to conventional 3D gravity and provides a transparent geometric interpretation of the double-copy construction. We relate the resulting theory to a Chern–Simons–like action, propose a higher-dimensional origin, and explore extensions that give rise to $AdS_3$
  solutions.
\end{abstract}

\maketitle


\section{Introduction}

Three-dimensional gravity has long attracted considerable attention, as it provides a theory of gravity without local propagating degrees of freedom and is therefore amenable to extensive analytical treatment. A major breakthrough was its reformulation as a Chern–Simons gauge theory \cite{Witten:1988hc}, which led to substantial progress in understanding its classical solution space, quantization, and boundary dynamics, as well as connections to topological field theory, knot invariants, and holography. For a recent review, see \cite{carlip2023quantumgravity21dimensions}.

In this note, we propose an alternative formulation of three-dimensional gravity in terms of vector frames. This reformulation is inspired by scattering amplitudes, in particular by the double-copy construction \cite{Bern:2008qj,Bern:2019prr} applied to three-dimensional Chern–Simons theory as proposed in \cite{Ben-Shahar:2021zww}. This theory (together with several related models) was studied within the Batalin–Vilkovisky (BV) formalism in \cite{Ben-Shahar:2024dju,Ben-Shahar:2025dci}. In the present work, we instead provide a simple analysis of the double-copy theory without any reference to the BV formalism with the aim of clarifying the gravitational nature of the double-copy construction.

The paper is organized as follows: in Section \ref{s:gravity} we review some basic facts about 3D gravity and introduce the relevant notation. In Section \ref{s:onshell} we reformulate the gravity theory as a theory of divergenceless vector frames and we show that it is on-shell equivalent to 3D gravity. Section \ref{s:double-copy} further elaborates on this theory and explains how it is related to the Chern-Simons-like action for the double copy theory on $\mathbb{R}^3$. 
In Section \ref{s:6D-origin} we offer a 6D interpretation of the double-copy theory and provide a simple derivation of many 3D properties from the 6D point of view. In Section \ref{s:ADS} we consider the non-local extension of our vector frame action which gives rise to AdS solutions in 3D. We finish with a summary and discussion of open questions in Section
\ref{s:summary}. 

\section{3D Gravity}\label{s:gravity}

In this section we review basic facts about 3D gravity and establish notation. Let us start with the standard Einstein-Hilbert (EH) action 
\begin{equation}
    S_{EH} = \int\limits_M d^3 x~\sqrt{g} ~R(g)~,
\end{equation}
which gives rise to the equations of motion $R_{\mu\nu}=0$. In 3D these imply $R_{\mu\nu\rho\sigma}=0$, so the metric $g$ is flat. The EH action is equivalent to the first-order action
\begin{equation}\label{BF3d-gravity}
  S = \int   e^a (d\omega_a +\frac{1}{2} \epsilon_{abc} \omega^b \wedge \omega^c )~,  
\end{equation}
provided that $g_{\mu\nu}= e^a_\mu \eta_{ab} e^b_\nu$ and $\det e \neq 0$. Here $e^a$ is the one-form vielbein and 
$\omega^a$ is the one-form spin connection. For the sake of clarity we assume Euclidean signature, and the generalization to Minkowski is straightforward. Here we use Latin letters for flat indices and Greek letters for curved indices. 
The action (\ref{BF3d-gravity}) implies the following equations 
of motion
\begin{eqnarray}
    de^a + \epsilon^a_{~~bc} e^b \wedge \omega^c=0~,\\
    d\omega^a + \frac{1}{2}\epsilon^a_{~~bc} \omega^b \wedge \omega^c=0~,
\end{eqnarray}
and we are interested in solutions with the property $\det e \neq 0$. The second equation is the flatness condition for the spin connection $\omega$ and it can be solved by  $\omega=0$ in every patch with constant transition functions between different patches. Thus the two equations collapse to one simple equation
\begin{equation}\label{eq-e-flat}
    d e^a=0~,
\end{equation}
where on the intersections of different patches $e^a$ can be glued by constant rotations. We are interested in those $e$ such that $\det e \neq 0$. 
Next let us introduce the dual vector frames satisfying
\begin{equation}
    E^\mu_a e_\mu^b = \delta^b_a~,~~~~~~ E^\mu_a e^a_\nu = \delta^\mu_\nu~.
\end{equation}
Assuming $\det e \neq 0$, the simple equation (\ref{eq-e-flat})  implies the following two equations for $E$
\begin{eqnarray}
    \{ E_a, E_b \}=0~, \label{eq-E-1}\\
    \partial_\mu \Big ( (\det E)^{-1} E^\mu_a \Big )=0~,\label{eq-E-constr}
\end{eqnarray}
where $\{~,~\}$ is the Lie bracket of vector fields. The first equation is the consequence of the identity
\begin{equation}
    \{ E_a, E_b\}^\nu e_\lambda^a e_\gamma^b = E^\nu_a (\partial_\gamma 
    e^a_\lambda - \partial_\lambda e^a_\gamma)~.
\end{equation}
The second equation is the consequence of the algebraic identities
\begin{eqnarray}
    \epsilon_{\mu\nu\rho} \det e = \epsilon_{abc}~ e^a_\mu e^b_\nu e^c_\rho~,\label{1-eq-forE}\\
    (\det e) E^\mu_a = \frac{1}{2}  \epsilon^{\mu\nu\rho} \epsilon_{abc} e^b_\nu e^c_\rho~,\label{2-eq-forE}
\end{eqnarray}
and $de^a=0$. To summarize, we have reduced the equations of motion of the EH action to the equations for the vector frames  (\ref{eq-E-1}) with the condition (\ref{eq-E-constr}) meaning that the frames are divergenceless
with respect to the volume form defined by the metric $g$. These equations allow us to glue the frames by constant rotations on the intersections and thus we do not lose any global information in our rewriting of the equations of motion. 
    
It will be convenient to compare to the Unimodular or partially gauge-fixed Henneaux-Teitelboim (HT) formulation of gravity \cite{Henneaux:1989zc,Buchmuller:1988wx,Buchmuller:1988yn}. In its covariant form HT contains an auxiliary vector density and is invariant under full diffeomorphisms. In fixed-volume gauge, where the divergence of that vector density is set equal to a background volume form, the action becomes
\begin{equation}
   S_{HT} = \int\limits_M  d^3x \Big (\sqrt{g} ~R(g) + \lambda (\sqrt{g} - \rho(x)) \Big )~,
\end{equation}
so the gauge symmetry is reduced to volume-preserving diffeomorphisms. In the first-order formalism we have   
\begin{align}
    S_{HT} = \int  \Big [&  e^a (d\omega_a +\frac{1}{2} \epsilon_{abc} \omega^b \wedge \omega^c ) \nonumber \\
    &+ \lambda (\frac{1}{6}\epsilon_{abc} e^a \wedge e^b \wedge e^c - \Omega ) \Big ]~, 
\end{align}
and the equations of motion from this action are
\begin{align}
    d\omega^a + \frac{1}{2} \epsilon^a_{~~bc} \omega^b \wedge \omega^c + \frac{\lambda}{2} \epsilon^a_{~~bc} e^b \wedge e^c=0~,\label{eq:HT-curv}\\
    d e^a + \epsilon^a_{~~bc} \omega^b \wedge e^c=0~,\label{eq:HT-tor}\\
    \frac{1}{6}\epsilon_{abc} e^a \wedge e^b \wedge e^c - \Omega =0~.\label{eq:HT-vol}
\end{align}
The Bianchi identities imply that $\lambda$ is constant. If we choose this constant to be zero then we go back to EH theory but with a fixed volume form. If $\lambda \neq 0$ then we go to the theory with non-zero cosmological constant and a fixed volume form. Depending on the context, we can regard fixing a volume form as a partial gauge fixing of the original theory, which is invariant under full diffeomorphisms. 
 
\section{On-shell equivalent theories}\label{s:onshell}

Now our goal is to define an action for vector frames that gives rise to the same equations of motion as in the previous section. Let us fix a volume form 
\begin{equation}
    \Omega= \rho(x) dx^1 \wedge dx^2 \wedge dx^3 = \frac{1}{6} \rho(x) \epsilon_{\mu\nu\rho} dx^\mu \wedge dx^\nu \wedge dx^\rho~.
\end{equation}
We also fix a flat structure, {\it i.e.}\ a trivialization of the tangent bundle with constant transition functions gluing $E_a^\mu$. We define the action
\begin{equation}\label{our-action}
    S_{DC} (E) = \frac{1}{6}\int d^3x~\rho^2 \epsilon^{abc} \epsilon_{\mu\nu\rho} E^\mu_a E^\nu_b E^\rho_c
\end{equation}
on the constrained frames $\partial_\mu (\rho  E^\mu_a)=0$ and we always assume $\det E \neq 0$. 
In coordinate-free fashion we write this action as follows
\begin{equation}\label{our-action-forms}
    S_{DC} =  \epsilon^{abc} \int \Big (\iota_{E_a}\iota_{E_b} \iota_{E_c} \Omega \Big ) \Omega~,
\end{equation}
and the constraint $\partial_\mu (\rho E^\mu_a)=0$ becomes
\begin{equation}
    d (\iota_{E_a}\Omega)=0~,
\end{equation}
which can be resolved locally as follows
\begin{equation}
    \iota_{E_a}\Omega = d\alpha_a
\end{equation}
for some locally defined 1-forms $\alpha_a$. 
On the space of divergenceless vector fields we can define the inner product   
\begin{align}
    &\langle v, w \rangle = \int \alpha (\iota_w \Omega) =
    \int \gamma (\iota_v \Omega) =\langle w, v \rangle~,\\
    &\hspace{2cm}\iota_v \Omega = d\alpha~,~~~\iota_w \Omega = d \gamma~, \nonumber
\end{align}
where one can check that despite the fact that $\alpha$ and $\gamma$ are locally defined, the inner product is globally defined (here we assume the absence of boundary terms). 
This inner product is invariant with respect to the Lie bracket of divergenceless vector fields. 

We can rewrite the action with this inner product.
First, using the Cartan calculus on differential forms, in particular 
\begin{equation}
    {\cal L}_X \iota_Y - \iota_Y {\cal L}_X = \iota_{\{X, Y \}}
\end{equation}
we have
\begin{equation}
    \epsilon^{abc} \int (\iota_{E_a}\iota_{E_b}\iota_{E_c} \Omega ) \Omega = 
    \epsilon^{abc} \int (\iota_{E_b}\iota_{E_c} \Omega ) \iota_{E_a}\Omega 
\end{equation}
and thus the action can be rewritten as follows
\begin{equation}\label{action-Lie}
    S_{DC} =  \epsilon^{abc} \langle E_a, \{ E_b , E_c \} \rangle~.
\end{equation}
If we vary this action over divergenceless vector fields we get the equation $\{ E_a, E_b \} =0$.  
Next let us discuss what is going on with the constraint on the solutions of equations of motion
\begin{equation}
    \partial_\mu (\rho E^\mu_a)=0~.
\end{equation}
Using the notation from the previous section we know that the equations $\{ E_a, E_b \}=0$ imply $de^a=0$ and therefore
\begin{equation}
    {\cal L}_{E_a} e^b = d \iota_{E_a} e^b + \iota_{E_a} d e^b =0~.
\end{equation}
If we rewrite the volume form $\Omega$ in the basis of $e$'s then  
\begin{equation}
    \Omega = \frac{\rho}{\det(e)} e^1 \wedge e^2 \wedge e^3 \ .
\end{equation}
Since our vector frames $E_a$ are divergenceless 
we have 
\begin{equation}
    {\cal L}_{E_a} \Omega =  d\iota_{E_a} \Omega =0~.
\end{equation}
Together with ${\cal L}_{E_a} e^b=0$, which is a consequence of the equation of motion, this implies that ${\cal L}_{E_a} \big(\rho/\det(e)\big) =0$. Thus we conclude that $\det e =  (\rm constant )~\rho$. Naturally, the action is constant on solutions of the equations of motion. Thus we reproduce the equations (\ref{eq:HT-curv})--(\ref{eq:HT-vol}) in the $\lambda=0$ sector.
The action (\ref{our-action-forms}) together with the constraint $\partial_\mu (\rho E_a^\mu)=0$ is invariant under volume-preserving diffeomorphisms 
\begin{equation}
    \delta E_a = {\cal L}_\xi E_a~,~~~~{\cal L}_\xi \Omega =0
\end{equation}
and constant frame rotations 
\begin{equation}
    E_a ~\rightarrow~\Lambda^b_a E_b~,
\end{equation}
where $\Lambda \in SO(3)$. Thus in flat trivialization we can glue locally defined $E$'s with constant rotations on the intersections of patches and thus it agrees with the discussion in the previous section. We do not lose any global information in this new description.

\section{Double copy theory}\label{s:double-copy}

 Let us study the action (\ref{our-action}) in more detail for $M={\mathbb R}^3$ and trivial flat structure. It is hard to implement the condition $\det E\neq 0$ explicitly. One way to deal with this problem is to study small perturbations. Let us choose the canonical flat volume form $\rho=1$ and assume the expansion
\begin{equation}\label{E=1+A}
    E_a^\mu = \delta^\mu_a + A^\mu_a~.
\end{equation}
If $A$ is small enough then we are guaranteed the condition $\det E \neq 0$. 
The presence of $\delta^\mu_a$ allows us to identify flat with curved indices. 
Algebraically we have the following identity
\begin{align}\label{identity}
    \frac{1}{6}\epsilon^{abc} \epsilon_{\mu\nu\rho} E_a^\mu E_b^\nu E_c^\rho &=\\
    1 + A^\rho_\rho + \frac{1}{2}& A^\nu_\nu A^\rho_\rho - \frac{1}{2} A^\rho_\nu A^\nu_\rho 
    + \frac{1}{6}\epsilon^{abc} \epsilon_{\mu\nu\rho} A_a^\mu A_b^\nu A_c^\rho~.\nonumber
\end{align}
Recall that our vector frame is divergenceless $\partial_\mu E^\mu_a=0$ and thus $\partial_\mu A^\mu_a=0$. This constraint together with standard identities for the Levi-Civita symbols
 allows us to rewrite the quadratic term as
\begin{equation}
    \frac{1}{2} \epsilon^{abc} \epsilon_{\mu\nu\rho}
    A_a^\mu \frac{\partial_b \partial^\nu}{\Box} A_c^\rho 
    = \frac{1}{2} A^\nu_\nu A^\rho_\rho - \frac{1}{2} A^\rho_\nu A^\nu_\rho
    \ ,
\end{equation}
where we need to restrict to such $A$'s that $\Box = \partial_\mu \partial^\mu$ is  invertible.
Now, substituting the above expression into the action (\ref{our-action}) with $\rho=1$, we get 
\begin{equation}\label{Maor-action}
    S_{DC} = \int d^3x \Big (  \frac{1}{2}
    A_a^\mu \frac{\partial_b \partial^\nu}{\Box} A_c^\rho +
    \frac{1}{6}   A_a^\mu A_b^\nu A_c^\rho  \Big )  \epsilon^{abc} \epsilon_{\mu\nu\rho} ~,
\end{equation}
where we removed the first term since it is $A$-independent and the second term since it is a total derivative under the integral. This is exactly the action that was derived from the double-copy procedure for Chern-Simons theory \cite{Ben-Shahar:2021zww}. If we vary this action over $A$'s satisfying $\partial_\mu A^\mu_a=0$ we obtain the equation 
\begin{equation}\label{eq-for-A}
    \partial_{[\mu} A_{\nu]}^\rho + A_{[\mu}^\lambda \partial_\lambda A_{\nu]}^\rho=0~,
\end{equation}
which is equivalent to (\ref{eq-E-1}) upon substitution (\ref{E=1+A}), as expected.
This theory can be regarded as a real version of Kodaira-Spencer (KS) gravity \cite{Bershadsky:1993cx}, for such a connection see \cite{Bonezzi:2024dlv} and for a recent discussion of the relation between 3D gravity and KS gravity see also \cite{Erdmenger:2025lvv}. Interestingly, this equation of motion can be seen as the vanishing of a field strength in the double copy of the non-linear sigma model with Yang-Mills theory, for this construction see \cite{Cheung:2021zvb,Kim:2025eqd}.
The action (\ref{Maor-action}) first appeared in \cite{Ben-Shahar:2021zww} in the context of double copy theory for the Chern-Simons theory and its similarity to KS gravity was discussed in \cite{Bonezzi:2024dlv,Ben-Shahar:2025dci}. Combining the present discussion with Section \ref{s:gravity}, equation (\ref{eq-for-A}) describes the fluctuations of the flat metric $g = \eta + \delta g$ around the canonical flat metric $\eta$ with fixed volume form
\begin{equation}
    R_{\mu\nu}(\eta + \delta g)=0~,~~~~\det (\eta + \delta g)=1~. 
\end{equation}
The relation between $g$ and $A$ is non-linear.   
 
In the double copy of Chern-Simons theory the physical fields also include $C^{\mu\nu}$ solving $\partial_\mu C^{\mu\nu}=0$ and $b_{ac}$ with action
\begin{align}
    S = 
    \int d^3 x\Big(&
    \frac{1}{2}A^\mu_a \frac{\partial^\nu \partial_b}{\Box}A^\rho_c - b_{ab}\frac{\partial_c \partial^\mu}{\Box}C^{\nu\rho} \nonumber \\
    &
    +\frac{1}{6}A^\mu_a A^\nu_b A^\rho_c - A^\mu_a b_{bc}C^{\nu\rho}
    \Big)\epsilon^{abc}\epsilon_{\mu\nu\rho} \ .
\end{align}
This is very natural from the point of view of deformation theory, see \cite{Ben-Shahar:2025dci} for details. The action 
corresponds to the physical part of the superfield actions \cite{Ben-Shahar:2021zww,Bonezzi:2024dlv,Ben-Shahar:2025dci}.

Using the identification (\ref{E=1+A}) and similar algebraic tricks discussed at the beginning of this section, we can rewrite this action as follows
\begin{equation}\label{3D-all-fields-gen}
    S= \int d^3 x \rho^2(x) \Big(
    \frac{1}{6}E^\mu_a E^\nu_b E^\rho_c - E^\mu_a b_{bc}C^{\nu\rho}
    \Big)\epsilon^{abc}\epsilon_{\mu\nu\rho}
\end{equation}
where we assume 
\begin{equation}
    \partial_\mu (\rho E^\mu_a) = 0 \ , \hspace{1cm} 
    \partial_\mu (\rho C^{\mu\nu}) = 0 \ ,
\end{equation}
 and we have allowed a general volume form. 
This condition is necessary in order to construct the action \cite{Bonezzi:2024dlv,Ben-Shahar:2025dci}, however there is still a gauge symmetry in the theory with transformations
\begin{eqnarray}
\label{double_copy_gauge_transf_2}
\delta E_a^\nu &=&  \xi^\lambda \partial_\lambda E^\nu_a - E^\lambda_a  \partial_\lambda \xi^\nu - C^{\nu\lambda}\partial_\lambda \beta_a ~, \cr
\delta b_{ab}&=&   \xi^\lambda\partial_\lambda b_{ab}+       ( E_a^\lambda \partial_\lambda\beta_b -E_b^\lambda\partial_\lambda\beta_a)~, \cr
\delta C^{\mu\nu}  &=&  \xi^\lambda \partial_\lambda C^{\mu\nu} -  C^{\lambda \nu }   \partial_\lambda \xi^\mu - C^{\mu\lambda} \partial_\lambda \xi^\nu  ~.
\end{eqnarray}
where $\xi^\nu$ and $\beta_a$ are two gauge parameters, and $\xi^\mu$ satisfies $\partial_\mu (\rho \xi^\mu )=0$. From the above action we can derive the equations of motion (for the case $\rho=1$ they are written explicitly in  \cite{Ben-Shahar:2025dci}), in Section \ref{s:6D-origin} we present a compact form for these equations of motion.

\section{6D origin of 3D gravity theory}\label{s:6D-origin}

In this section we want to explain that all formulas presented in the previous section have a 6D origin and both the action and symmetries follow easily from the 6D perspective. 

We start by recalling the Cartan calculus for multivector fields \cite{marle2008calculusliealgebroidslie}. The simplest way is to extend the normal Cartan calculus from vector fields to multivector fields. Let us consider the multivector field of the form
\begin{equation}
    X= X_1 \wedge ... \wedge X_k 
\end{equation}
 such that each $X_i$ has degree $|X_i|=1$. We define the operation of contraction of a multivector field $X$ with a differential form $\omega$ as follows
\begin{equation}
    \iota_{X_1 \wedge ... \wedge X_k} \omega = \iota_{X_k} (\iota_{X_1 \wedge ... \wedge X_{k-1}} \omega)~,
\end{equation}
and it can be extended to all multivector fields by linearity. 
Next define the Lie derivative with respect 
to multivector field $X$ acting on differential forms as
\begin{equation}
    {\cal L}_X = d\iota_X - (-1)^{|X|} \iota_X d
\end{equation}
where the degree of the generalized Lie derivative is $|{\cal L}_X|=|X|-1$.
We have the following identities for graded commutators
\begin{equation}
     [{\cal L}_P, {\cal L}_Q] = {\cal L}_{\{P, Q\}}~, 
\end{equation}
and 
\begin{equation}\label{Cartan-multi-1}
    [   {\cal L}_P, \iota_Q ]  =  \iota_{\{P, Q\}}
\end{equation}
where $\{P, Q\}$ is the Schouten bracket of multivector fields which is the extension of the Lie bracket to multivector fields. If we fix a volume form $\Omega$ then the multivector field $X$ is divergenceless if 
\begin{equation}
     d (\iota_X \Omega)=0~.
\end{equation}

Let us specialize to 6D and choose a fixed volume form $\Omega_6$. Then we can define the action 
\begin{equation}\label{PG-action1}
    S_{PG} = \int (\iota_\pi \iota_\pi \iota_\pi \Omega_6) \Omega_6
\end{equation}
over bivector fields $\pi$ which are divergenceless with respect to  $\Omega_6$
 \begin{equation}
     d (\iota_\pi \Omega_6)=0~. 
 \end{equation}
 In local coordinates we can write this as follows
\begin{equation}\label{6D-indices}
    S_{PG} = \frac{1}{6!} \int d^6 y ~\rho^2(y)~ \pi^{MN} \pi^{LK} \pi^{SR} \epsilon_{MNLKSR}
\end{equation}
with the constraint on the fields
\begin{equation}
    \partial_M(\rho \pi^{MN} )=0~. 
\end{equation}
The action (\ref{PG-action1}) can be rewritten using a pairing between divergenceless multivector fields $P$ and $Q$ defined as follows
\begin{eqnarray}
    \langle P, Q \rangle = \int \iota_P \Omega_6 \wedge \alpha_Q~,
\end{eqnarray}
where 
\begin{equation}
    \iota_Q \Omega_6 = d\alpha_Q~.
\end{equation}
Here $\alpha_Q$ is locally defined but one can check that the pairing is well-defined and it does not depend on the choice of $\alpha_Q$. The pairing is non-zero if $|P|+|Q|=5$. 

This pairing is invariant with respect to the Schouten bracket of divergenceless multivector fields,
\begin{equation}
    \langle \{P, Q\} , R\rangle  = -\langle P, \{Q , R\}\rangle  
\end{equation}
Thus the action $S_{PG}$ can be written as
\begin{equation}\label{PG-action2}
    S_{PG} = \langle \pi, \{ \pi, \pi \} \rangle~,
\end{equation}
where we have used the identity (\ref{Cartan-multi-1}).
If we vary this action over divergenceless bivector fields then we get the Poisson condition
\begin{equation}
    \{\pi, \pi\}=0~,
\end{equation}
where for variation we used 
\begin{equation}
    \iota_\pi \Omega_6 = d\alpha~. 
\end{equation}
Thus the action (\ref{PG-action2}) is extremized on unimodular (divergenceless) Poisson structures.  
The action (\ref{PG-action2}) is invariant under volume-preserving diffeomorphisms which can be written simply as
\begin{equation}\label{6D-diffs}
    \delta \pi = {\cal L}_\Xi \pi = \{ \Xi, \pi \}~,
\end{equation}
where $\Xi$ is a 6D vector field which preserves  $\Omega_6$.

Now let us consider a particular six-dimensional manifold, $M_6 = M_3 \times T^3$ with coordinates $(x, \phi)$ correspondingly. Let us consider a volume form $\Omega_6$ 
   
\begin{equation}
    \epsilon_{abc} \rho (x) dx^1 \wedge dx^2 \wedge dx^3 \wedge d\phi^a \wedge d\phi^b \wedge d\phi^c
\end{equation}
    and a bivector $\pi$ 
\begin{equation}\label{6D-PS-ansatz}
    \pi = E^\mu_a (x) \partial_\mu \wedge \partial^a + b_{ab}(x) \partial^a \wedge \partial^b + C^{\mu\nu}(x) \partial_\mu \wedge \partial_\nu   
\end{equation}
which are independent of the torus directions  $\phi$. Here we use the flat metric $\eta_{ab}$ to raise and lower indices in the $\phi$ directions. 
If we evaluate the 6D action (\ref{6D-indices}) on such $\Omega_6$ and $\pi$ then we get the 3D action (\ref{3D-all-fields-gen}) from the previous section (up to an overall sign). 
The 6D equations of motion  $\{\pi, \pi\}=0$ for (\ref{6D-PS-ansatz}) imply the following
\begin{eqnarray}
       \epsilon^{abc} {\cal L}_{E_a} b_{bc}=0~,\\
       {\cal L}_{E_a} C=0~,\\
       \{E_a, E_b \} + 4 \{C, b_{ab}\}=0~,
\end{eqnarray}
while the condition 
\begin{equation}
    \{ C, C\}=0
\end{equation}
works automatically in 3D due to $\partial_\mu (\rho C^{\mu\nu})=0$. This is a compact form of the equations of motion which can be derived from the 3D action (\ref{3D-all-fields-gen}). If  in the 6D symmetries (\ref{6D-diffs}) we take $\Xi^M \partial_M = \xi^\mu \partial_\mu + \beta_a \partial^a$ to be independent of $\phi$-directions then the divergenceless condition becomes $\partial_\mu(\rho \xi^\mu) = 0$ and we get 3D transformations
\begin{eqnarray}
    \delta E_a = {\cal L}_\xi E_a - \{C, \beta_a\}~,\\
    \delta b_{ab} = {\cal L}_\xi b_{ab} + {\cal L}_{E_a} \beta_b - {\cal L}_{E_b} \beta_a ~,  \\
    \delta C = {\cal L}_\xi C~,
\end{eqnarray}
where we use 3D notation for the Lie derivative and the Schouten bracket. These transformations coincide with the gauge transformations (\ref{double_copy_gauge_transf_2}) from the previous section. 
  
\section{Non-local action and AdS solution}\label{s:ADS}

So far we have studied the local action (\ref{action-Lie}) and its possible local 
extensions. However, in 3D there exists a quadratic non-local term which can be added to the action   (\ref{action-Lie}). Let us define the quadratic term through the pairing of divergenceless vector fields defined in Section \ref{s:onshell}
\begin{equation}\label{AdS-term}
   \langle E^a, E_{a} \rangle = \int \alpha^a \wedge \iota_{E_a} \Omega ~, 
\end{equation}
here $\iota_{E_a}\Omega = d \alpha_a$ and we contract flat indices using the flat metric $\eta_{ab}$ and  this term is non-local. 
 If we assume a flat background metric and canonical flat $\Omega$ then up to normalization factors we can rewrite this term as follows 
\begin{equation}
\int d^3x~E^{a\mu} \frac{\partial^\nu}{\Box} E_a^\rho \epsilon_{\mu\nu\rho}~.
\end{equation} 
  It is important to stress that the definition (\ref{AdS-term}) requires only a background volume form. 
Let us write a new action with the addition of this term
 \begin{equation}
 S_{DC}(E)=  \frac{\gamma}{2}  \langle E^a, E_a \rangle + \frac{1}{6} \epsilon^{abc} \langle E_a, \{ E_b, E_c \} \rangle~,   
 \end{equation}
 where $\gamma$ is a parameter. If we vary this action over divergenceless vector fields we get the following equation 
 \begin{equation}
      \epsilon^{abc} \{E_b, E_c \} = 2 \gamma  E^a~,
\end{equation}
which is equivalent to
\begin{equation}
    \{ E_a, E_b \} = \epsilon_{ab}^{~~~c} \gamma E_c~,
\end{equation}
where for definiteness we used the Euclidean signature. This equation implies the following conditions on $e$'s
\begin{equation}\label{MC-AdS}
     de^a + \gamma  ~ \epsilon_{bc}^{~~~a} e^b \wedge e^c =0~,
\end{equation}
which can be mapped to solutions of 3D gravity with negative cosmological constant.
   
Let us recall that the first-order action of $3d$-gravity with cosmological constant $\Lambda$ has the form
\begin{equation}\label{BF3d-gravity-CT}
  S = \int  \Big [  e^a (d\omega_a +\frac{1}{2} \epsilon_{abc} \omega^b \wedge \omega^c ) - \frac{\Lambda}{6} \epsilon_{abc} e^a \wedge e^b \wedge e^c \Big ]~,  
\end{equation}
and it gives rise to the following equations of motion 
\begin{eqnarray}
    de^a + \epsilon^a_{~~bc} e^b \wedge \omega^c=0~,\label{CC-1}\\
    d\omega^a + \frac{1}{2}\epsilon^a_{~~bc} \omega^b \wedge \omega^c= \frac{\Lambda}{2} \epsilon^{a}_{~~bc} e^b \wedge e^c   ~.\label{CC-2}
\end{eqnarray}
These equations can be reduced to (\ref{MC-AdS}) if we make the following choices
\begin{equation}\label{cond-ads}
     \omega^a =\gamma e^a~,~~~~~-\Lambda = \gamma^2~. 
\end{equation}
 Thus for $\Lambda <0$ we can choose real $\gamma= \pm\sqrt{|\Lambda|}$ and let us fix the plus sign in what follows. 
Here we are solving the equations of motion  (\ref{CC-1}) and (\ref{CC-2}) in a particular gauge with respect to $SO(3)$ gauge transformations. Following  \cite{Witten:1988hc} the equations 
  (\ref{CC-1}) and (\ref{CC-2}) are equivalent to the flatness of two connections $\omega^a \pm \sqrt{|\Lambda|}e^a$ and in 
   the condition (\ref{cond-ads}) we can set only one of them to zero since we need to require $\det e \neq 0$. 
Since $\det e \neq 0$ we have the algebraic relations (\ref{1-eq-forE}) and (\ref{2-eq-forE}), thus using these algebraic relations and the equation (\ref{MC-AdS}) we can show that $E_a$ are divergenceless with respect to $\det e$
\begin{equation}\label{AdS-div}
    \partial_\mu (\det e ~E^\mu_a)=0~. 
\end{equation}
Therefore the $AdS_3$ metric can be encoded into the equations (\ref{AdS-div}) and (\ref{MC-AdS}) with $\gamma= \sqrt{|\Lambda|}$ for the vector frames. 

Thus to summarize: upon choice of a background volume form $\Omega$ we can define the non-local action 
\begin{equation}
 S_{DC}(E)=  \frac{\sqrt{|\Lambda|}}{2}  \langle E^a, E_a \rangle + \frac{1}{6} \epsilon^{abc} \langle E_a, \{ E_b, E_c \} \rangle~,   
\end{equation}
where $E$ are divergenceless vector fields with respect to $\Omega$. This action is extremized on $E$'s which satisfy $\{E_a, E_b\} = \sqrt{|\Lambda|}\epsilon_{ab}^{~~~c} E_c$. Moreover, on-shell $\Omega$ is equal to $(\det e) dx^1 \wedge dx^2 \wedge dx^3$ up to an overall constant, and the solutions to these equations can be mapped to solutions to equations \eqref{eq:HT-curv}-\eqref{eq:HT-vol}.

\section{Conclusion}\label{s:summary}

In this note we have presented a simple motivation and derivation for the physical part of the double copy theory of Chern-Simons theory. This discussion clarifies the relation of the double copy theory to gravity in simple terms in the context of 3D gravity. This work is complementary to our previous work \cite{Ben-Shahar:2025dci}, but here we avoided any reference to the BV formalism and other formal constructions. Some of these ideas have a straightforward generalization to higher-dimensional topological gravities. 

The fact that 3D gravity can be recast as a real analog of the Kodaira-Spencer theory with a 6D origin suggests that the double copy theory is closely related to string field theory of some 
2D model (e.g., 2D Poisson sigma model with 6D target). We leave the details of this observation for a future publication.

\begin{acknowledgments}
We thank Joonhwi Kim for useful discussions and comments on an earlier draft. The research of M. B. S. is supported by the Knut and Alice
Wallenberg Foundation (grant KAW 2023.0490). The research of M.\ Z.\ is supported by the Swedish Research Council excellence center grant ``Geometry and Physics'' 2022-06593.
 
\end{acknowledgments}


\bibliography{3d_gravity}

\end{document}